\documentclass[runningheads]{llncs}
\usepackage[T1]{fontenc}
\usepackage{amsmath}
\usepackage{multicol}
\usepackage{amsfonts}
\usepackage{xcolor}
\usepackage{graphicx}
\usepackage{tcolorbox}
\usepackage{makecell} 
\usepackage{float}

\usepackage{tcolorbox}
\usepackage{tikz}
\usepackage{algorithm}
\usepackage{algpseudocode}
\usepackage{caption}
\usepackage{subcaption}
\usepackage{cite}

\usepackage{url}
\usepackage{hyperref}
\hypersetup{
    colorlinks=true,
    linkcolor=blue,
    filecolor=magenta,      
    urlcolor=cyan,
}
\begin{document}
\title{A Reverse Reachable Set Based Approach for Motif Oriented Profit maximization in Social Networks}
\titlerunning{Motif Oriented Profit maximization in Social Networks}
\author{Poonam Sharma \and Suman Banerjee }
\authorrunning{Sharma and Banerjee} % abbreviated author list (for running head)
\institute{Indian Institute of Technology Jammu,
J \& K-181221, India. \\
\email{\{poonam.sharma,suman.banerjee\}@iitjammu.ac.in}}
\maketitle
\begin{abstract}
 Profit Maximization is one of the key objectives for social media marketing, where the task is to choose a limited number of highly influential nodes such that their initial activation leads to maximum profit. In this paper, we introduce a variant of the Profit Maximization Problem where we consider that instead of nodes, benefits are assigned to some of the motifs of the graph, and these benefit values can be earned once a given threshold count of nodes from the motifs is influenced. The goal here is to choose a limited number of nodes for initial activation (called `seed nodes') such that the motif-oriented profit gets maximized. Formally, we call our problem the \textsc{Motif Oriented Profit Maximization} Problem. We show that the problem is NP-hard to solve optimally. We propose a Reverse Reachable Set-based framework to solve our problem. The proposed methodology broadly divides into three steps: KPT Estimation and $\mathcal{RR}$ Set generation, Seed Set Selection, and Motif Oriented Profit Estimation. The proposed methodology has been analyzed to understand its time and space requirements. It has been implemented with real-world social network datasets, and the results are reported. We observe that the seed set selected by the proposed solution approaches leads to more profit compared to the seed sets selected by the existing methods. The whole implementation and data are available at: https://github.com/PoonamSharma-PY/MotifProfit.
\end{abstract}
\section{Introduction}\label{Sec:Intro}
In recent times, \emph{Online Social Networks} play a pivotal role in spreading news, ideas, rumors, etc., and this happens due to the diffusion of information \cite{bakshy2012role,arnaboldi2017online}. People tend to share the information through social media posts, and people who are in the friend (or follower) list may wish to like, share, comment, etc. on the post. If the person is influential, then there is a very high chance that many of the users of the network will come to know about the fact. This phenomenon has been exploited by commercial houses for promoting their brands. For this purpose, they choose a limited number of influential people from the network and distribute free (or discounted) products with the hope that they will spread positive words about the product due to word-of-mouth. This notion is called Viral Marketing through Social Media. In recent times, commercial houses spend a significant portion of their revenue on Social Media advertisements.  
\par To study the diffusion process in a social network, several models have been introduced and studied in the literature. Among them, the Independent Cascade Model is the most popular one. In the context of viral marketing, the key computational problem that arises is that given a social network, how can we effectively select a limited number of nodes for initial activation such that the influence gets maximized? Initially, this problem was posed by Domingos and Richardson \cite{domingos2001mining,richardson2002mining}. Later, Kempe et al. \cite{kempe2003maximizing,kempe2005influential} showed that this problem is NP-hard to solve optimally under the IC Model of diffusion. They proposed an iterative greedy approach based on marginal influence gain computation, which provides a $(1-\frac{1}{e})$-factor approximate solution. This study triggers a significant amount of research in this direction, and a huge amount of literature is available. The proposed solution approaches for this problem can be classified into the following categories: Approximation Algorithms \cite{leskovec2007cost}, Heuristic Solutions \cite{chen2010scalable}, Soft Computing-based Approaches, Reverse Reachable Set-based Approaches, and many more. 
\par In practice, social networks are formed by rational human beings, which means if a user is acting as a seed user, then (s)he must be incentivized. Also, in commercial advertising, the key objective is to maximize profit. Hence, the following problem is of immense importance: Given a social network, the cost and benefit of each user, and a fixed budget, how can we select a seed set within the budget to maximize the profit? This problem has been referred to as the Profit Maximization Problem. In the past decade, this problem has been studied extensively in the literature. In most of the studies, it has been considered that every user of the network has some benefit value that can be earned if the user is influenced. Now, consider the following scenario. A group of friends wants to dine in a restaurant. Now, any restaurant brand will be able to attract this group and earn some profit if that brand can influence the whole group. A small group of nodes in a network is called a motif. Sometimes it is important to consider influencing a whole motif rather than an individual user. To influence a motif, it may be sufficient to influence the majority of the users present in the motif. In this paper, we consider the problem of maximizing the profit of a commercial campaign by influencing the motifs. We call this problem the \textsc{Motif Oriented Profit Maximization} Problem. In this problem, we are given a social network where each user is assigned a selection cost, a set of motifs along with their corresponding benefit value, and a budget. This problem asks to choose a subset of nodes within the allocated budget such that the earned profit by maximizing the influence among the motifs is maximized. To the best of our knowledge, we are the first to study the Profit Maximization Problem under the motif-oriented setup. In particular, we make the following contributions in this paper: 
\begin{itemize}
    \item We introduce and study the \textsc{Motif Oriented Profit Maximization} Problem for which there does not exist any literature.
    \item We propose a reverse reachable set-based solution approach to solve our problem with a detailed analysis and illustration.
    \item  A number of experiments have been conducted on real-world social network datasets, and the results are compared with the existing methods to show the effectiveness and efficiency of the proposed solution approach.
\end{itemize}
The rest of the paper has been organized as follows. Section \ref{Sec:Problem} describes background information and defines the problem formally. Section \ref{Sec:Solution} describes the proposed solution approaches with a detailed analysis. The experimental evaluation of the proposed solution approaches has been described in Section \ref{Sec:Experiments}. Finally, Section \ref{Sec:Conclusion} concludes our study and gives future research directions.

\section{Background and Problem Definition} \label{Sec:Problem}
In this section, we describe the required preliminary concepts and subsequently define our problem formally. Initially, we start by describing the notion of social networks.
\subsection{Social Networks}
A social network is defined as an interconnected structure among a group of people, which has been formally stated in Definition \ref{Def:SN}.
\begin{definition}[Social Networks] \label{Def:SN}
A social network is often represented as a simple, (un)directed, weighted graph $\mathcal{G}(\mathcal{V}, \mathcal{E}, \mathcal{P})$ where the vertex set $\mathcal{V}$ represents the set of users connected through the network, the edge set represents the social relationships, and the edge weight function $\mathcal{P}$ maps each edge to its corresponding influence probability, i.e., $\mathcal{P}: \mathcal{E} \longrightarrow (0,1]$.
\end{definition}
For any edge $(u_iu_j) \in \mathcal{E}$, it means that $u_i$ and $u_j$ are in the social relationship. We reserve $n$ and $m$ to denote the number of nodes and edges, respectively. For any edge $(u_iu_j) \in \mathcal{E}$, its influence probability is denoted by $\mathcal{P}(u_iu_j)$. In our study, we assume that every edge has a non-zero influence probability, i.e., for any edge $(u_iu_j) \in \mathcal{E}$, $0 < \mathcal{P}(u_iu_j) \leq 1$.
\par In graph data analytics, there has always been an interest in understanding how large networks (e.g., Social Networks, Biological Networks, etc.) have been formed. It has been found that a large network is formed using small networks as a building block, which is also called a Motif, as stated in Definition \ref{Def:Motif}.
\begin{definition}[Motif] \label{Def:Motif}
A motif is a subgraph that appears significantly more often in a real network than would be expected in a randomized network with the same number of nodes and edges.
\end{definition} 
\subsection{Information Diffusion and Social Influence Maximization}
Among many, one of the properties of social networks is the diffusion of information, which says that an individual connected through online social networks tends to share the information. Now, it is expected that if the person is influential, then he will have many social neighbors, and a large number of them will be influenced by the information and share it further. This process will be continued, and the hope is that at the end of the diffusion process, a large number of people will be influenced. The entire process is referred to as the \emph{Information Diffusion}. This process starts from a set of initially active nodes referred to as Seed Nodes. How the information diffusion happens in the network depends on the diffusion model that has been chosen. In this study, we assume that the diffusion of information is happening by the rule of the IC Model, which has been stated in Definition \ref{Def:IC}.
\begin{definition}[Independent Cascade Model] \label{Def:IC}
As per the IC Model, the diffusion process starts from a set of initially active nodes called seed nodes and proceeds in discrete time steps. In the diffusion process, an active node at time step $t$ will get a single chance to activate its inactive neighbors. A node's state can be either `activated' (also known as influenced) or `non-activated' (also called non-influenced). A node can change its state from `non-activated' to `activated'; however, it cannot do so vice versa. The diffusion process stops when no more node activation is possible.
\end{definition} 
At the end of the diffusion process, the number of influenced nodes is called the influence of the seed set. For any given seed set $\mathcal{S} \subseteq \mathcal{V}$, $I(\mathcal{S})$ denotes the set of influenced nodes and $\sigma( \mathcal{S})$ denotes the influence of $\mathcal{S}$, where $\sigma()$ is the social influence function (a set function defined on the ground set $\mathcal{V}$) which maps each subset of the users of the network to their expected influence, i.e., $\sigma: 2^{\mathcal{V}} \longrightarrow \mathbb{R}^{+}_{0}$. In the IC Model, for a given seed set, its influence can be computed by constructing $2^{m}$ many live graphs and taking the expected value as described in \cite{kempe2003maximizing}. As mentioned in the literature, the influence under the IC Model of diffusion is non-negative, monotone, and sub-modular. In the context of information diffusion, one well-studied problem is the Social Influence Maximization Problem which has been stated in Definition \ref{Def:SIM}.
\begin{definition}[Social Influence Maximization Problem] \label{Def:SIM}
Given a social network $\mathcal{G}(\mathcal{V}, \mathcal{E}, \mathcal{P})$, and a positive integer $k$, the problem of social influence maximization asks to choose $k$ many users to activate initially, such that the maximum number of nodes gets influenced at the end of the diffusion process. Mathematically, this problem can be posed as an optimization problem as mentioned in Equation \ref{Eq:problem}.
\begin{equation} \label{Eq:problem}
    \mathcal{S}^{OPT} \longleftarrow \underset{\mathcal{S} \subseteq \mathcal{V} \text{ and } |\mathcal{S}|=k}{argmax} \ \sigma(\mathcal{S})
\end{equation}
\end{definition} 
\subsection{Profit Maximization in Social Networks}
In commercial campaigns, the users of the network need to be incentivized, and every user of the network is assigned some benefit value, which can be earned if the user is influenced. These notions have been formalized by the Cost and Benefit functions which are denoted by $\mathcal{C}$ and $b$, respectively, i.e.,  $\mathcal{C}: \mathcal{V} \longrightarrow \mathbb{R}^{+}$ and $b: \mathcal{V} \longrightarrow \mathbb{R}_{0}^{+}$. For any user $u \in \mathcal{V}$, its cost and benefit are denoted by $\mathcal{C}(u)$ and $b(u)$, respectively. Now, we define the notion of the earned profit by a seed set $\mathcal{S}$ in Definition \ref{Def:EP}.

\begin{definition}[Earned Profit] \label{Def:EP}
Given a seed set $\mathcal{S}$, the earned profit by $\mathcal{S}$ is defined as the difference between the earned benefit by the seed set and the cost of the seed set. This is denoted by $\Phi(\mathcal{S})$ and can be mathematically posed in Equation \ref{Eq:profit}.
\begin{equation} \label{Eq:profit}
    \Phi(\mathcal{S})= \underset{u \in I(\mathcal{S})}{\sum} \ b(u) - \underset{u \in \mathcal{S}}{\sum} \mathcal{C}(u)
\end{equation}
\end{definition} 
Naturally, in a commercial campaign, it is important to select a limited number of influential nodes within the budget to maximize the profit. This problem has been referred to as the Profit Maximization Problem in the literature and stated in the Definition \ref{Def:PMP}.
\begin{definition}[Profit Maximization Problem] \label{Def:PMP}
Given a social network, \\ $\mathcal{G}(\mathcal{V},\mathcal{E},\mathcal{P})$ with the Cost and Benefit functions $\mathcal{C}: \mathcal{V} \longrightarrow \mathbb{R}^{+}$ and $b: \mathcal{V} \longrightarrow \mathbb{R}_{0}^{+}$, respectively, and a fixed budget $\mathcal{B}$, the profit maximization problem asks to choose a set of nodes for initial activation such that the earned profit by the seed set is maximized. Mathematically, this problem can be posed as shown in Equation \ref{Eq:PMP}.
\begin{equation}\label{Eq:PMP}
    \mathcal{S}^{OPT} \longleftarrow \underset{\mathcal{S} \subseteq \mathcal{V} \text{ and } \underset{u \in \mathcal{S}}{\sum} \mathcal{C}(u) \leq \mathcal{B}}{argmax} \ \Phi(\mathcal{S})
\end{equation}
\end{definition} 
$ \mathcal{S}^{OPT}$ denotes the optimal seed set for the budget $\mathcal{B}$ in $\mathcal{G}$. As mentioned previously, this problem has been studied in the literature and a number of solution methodologies have been proposed. However, as mentioned in Section \ref{Sec:Intro}, we study the Motif Oriented Profit Maximization Problem. In this problem, we assume that along with the input social network $\mathcal{G}(\mathcal{V}, \mathcal{E}, \mathcal{P})$, we are also given a set of $\ell$ motifs $\mathcal{M}=\{m_1, m_2, \ldots, m_{\ell}\}$ and a benefit function $b$ that maps each of the motifs to its corresponding benefit value, i.e., $b: \mathcal{M} \longrightarrow \mathbb{R}_{0}^{+}$. For any arbitrary motif $m_j \in \mathcal{M}$, $|m_j|$ denotes the number of vertices that the motif contains. The associated benefit with the motif is denoted by $b(m_j)$. This benefit can be earned if the motif is influenced. Now, how do we decide whether a motif has been influenced or not? This depends on the influence model. In this study, we assume that a threshold has been given, and if at least the threshold number of nodes of the motif are influenced, for the motif $m_j \in \mathcal{M}$, its threshold is denoted by $\tau_j$, $1 \leq \tau_j \leq |m_j|$. Given a seed set $\mathcal{S} \subseteq \mathcal{V}$, for every motif $m_j \in \mathcal{M}$, we define an indicator boolean variable $I_{m_j}(\mathcal{S})$ which takes the value $1$ if the motif is influenced and $0$, otherwise. This has been mentioned in the Conditional Equation \ref{Eq:Motif}.
\begin{equation} \label{Eq:Motif}
  I_{m_j}(\mathcal{S}) =
  \begin{cases}
    1, & \text{if } |I(\mathcal{S}) \cap V(m_j)| \geq \tau_j \\
    0, & \text{ otherwise } 
  \end{cases}
\end{equation}
Now, we define the notion of Motif Oriented Earned Profit by a given seed set, which is stated in Definition \ref{Def:Motif_Profit}. 
\begin{definition}[Motif Oriented Earned Profit] \label{Def:Motif_Profit}
Given a Social Network $\mathcal{G}(\mathcal{V}, \mathcal{E}, \mathcal{P})$, a seed set $\mathcal{S}$, and a set of motifs $\mathcal{M}=\{m_1, m_2, \ldots, m_{\ell}\}$ with the corresponding benefit function $b: \mathcal{M} \longrightarrow \mathbb{R}_{0}^{+}$, and the cost function $\mathcal{C}: \mathcal{V} \longrightarrow \mathbb{R}^{+}$, the motif oriented earned profit by the seed set $\mathcal{S}$ is defined as the difference between the expected motif oriented earned benefit by the seed set and the cost of the seed set. This has been mathematically represented in Equation \ref{Eq:Motif_Profit}. 

\begin{equation} \label{Eq:Motif_Profit}
    \Phi_{\mathcal{M}}(\mathcal{S})= \underset{g \in L(\mathcal{G})}{\sum} \ Pr(g) \underset{m_j \in \mathcal{M}}{\sum} \ I_{m_j}(\mathcal{S}) \cdot b(m_j) - \underset{u \in \mathcal{S}}{\sum} \ \mathcal{C}(u)
\end{equation}
\end{definition} 
The following question arises: Given a seed set, how efficiently can we compute its motif-oriented earned profit? Theorem \ref{Th:Profit_Computation} states the fact.
\begin{theorem} \label{Th:Profit_Computation}
Given a Social Network $\mathcal{G}(\mathcal{V}, \mathcal{E}, \mathcal{P})$, a seed set $\mathcal{S}$, and a set of motifs $\mathcal{M}=\{m_1, m_2, \ldots, m_{\ell}\}$, accurately computing the motif oriented earned profit is a \#P-Complete Problem.
\end{theorem}
Now, in a commercial campaign, of course, it is important to choose the seed set effectively. Within the allocated budget, we formally state the Motif Oriented Profit Maximization Problem in Definition \ref{Def:Problem}.

\begin{definition}[Motif Oriented Profit Maximization Problem] \label{Def:Problem}
Given a Social Network $\mathcal{G}(\mathcal{V}, \mathcal{E}, \mathcal{P})$, a set of motifs $\mathcal{M}=\{m_1, m_2, \ldots, m_{\ell}\}$ with the corresponding benefit function $b: \mathcal{M} \longrightarrow \mathbb{R}_{0}^{+}$, cost function $\mathcal{C}: \mathcal{V} \longrightarrow \mathbb{R}^{+}$, and a fixed budget $\mathcal{B}$, this problem asks to choose a seed set to maximize the Motif Oriented Earned Profit as stated in Definition \ref{Def:Motif_Profit} such that the total cost of the seed set is less than the budget. Mathematically, this problem has been stated in Equation \ref{Eq:Profit}.
\begin{equation} \label{Eq:Profit}
    \mathcal{S}^{OPT} \longleftarrow \underset{\mathcal{S} \subseteq \mathcal{V} \text{ and } \underset{u \in \mathcal{S}}{\sum} \mathcal{C}(u) \leq \mathcal{B}}{argmax} \ \Phi_{\mathcal{M}}(\mathcal{S})
\end{equation}
\end{definition} 
As mentioned in \cite{lu2012profit}, the Profit Maximization Problem is NP-hard. Motif Oriented Profit Maximization Problem is a generalization of the Profit Maximization Problem; hence, Motif Oriented Profit Maximization Problem will also remain NP-hard. This has been formally stated in Theorem \ref{Th:Hardness}.
\begin{theorem} \label{Th:Hardness}
    Given a Social Network $\mathcal{G}(\mathcal{V}, \mathcal{E}, \mathcal{P})$, a set of motifs $\mathcal{M} = \{m_1, m_2, \\ \ldots, m_{\ell}\}$ with the corresponding benefit function $b: \mathcal{M} \longrightarrow \mathbb{R}_{0}^{+}$, cost function $\mathcal{C}: \mathcal{V} \longrightarrow \mathbb{R}^{+}$, and a fixed budget $\mathcal{B}$, finding an optimal seed set to maximize the Motif Oriented earned profit is NP-hard.
\end{theorem}
Next, we proceed to describe the solution methodologies subsequently.
\section{Proposed Approach} \label{Sec:Solution}
Our proposed solution approach is based on the notion of Reverse Reachable Set, which has been stated in Definition \ref{Def:RR_Set}.
\begin{definition} [Reverse Reachable Set] \cite{tang2014influence} \label{Def:RR_Set}
    Given a Social Network $\mathcal{G}(\mathcal{V}, \mathcal{E}, \mathcal{P})$, and a node $v$, the reverse reachable set of $v$ is denoted by $\mathcal{RR}(v)$ and defined as the set of nodes from which there exists a directed path to the node $v$ which has been stated in Equation $\ref{Eq:RRSet}$ 
\begin{equation}  \label{Eq:RRSet}
    \mathcal{RR}(v)=\{u: \text{ There exists a path from } u \text{ to } v\}
\end{equation}
\end{definition}
\vspace{-0.2in}
\begin{algorithm}[!htbp]
\caption{Motif Oriented RIS Framework}
\label{algorithm1:motif_ris}
\begin{algorithmic}[1]
\Statex \textbf{Input:} Graph $\mathcal{G}(\mathcal{V}, \mathcal{E}, \mathcal{P})$, Cost function $\mathcal{C}(\cdot)$, Benefit function $b(\cdot)$, Motif set $\mathcal{M}$, Budgets $\mathbb{B}$, Thresholds $\tau$, Simulation count $T$
\Statex \textbf{Output:} Seed set $\mathcal{S} \subseteq \mathcal{V}$
\For{each budget $\mathcal{B} \in \mathbb{B}$}
    \State $k \gets \left\lfloor \mathcal{B} / \min_v \mathcal{C}(v) \right\rfloor$
    \State $\kappa \gets \textsc{EstimateKPT}(\mathcal{G}, k, \mathcal{C}, b)$
    \State $\theta \gets \textsc{ComputeTheta}(\kappa, |V|, k)$
    \State $\mathcal{R} \gets \textsc{GenerateRRsets}(\theta, \mathcal{G})$
    \State $\mathcal{S} \gets \textsc{GreedySeedSelection}(\mathcal{R}, \mathcal{C}, b, \mathcal{B})$
    \State Perform $T$ Monte Carlo simulations of diffusion from $\mathcal{S}$ to obtain $\{A_1, A_2, \dots, A_T\}$
    \State Compute average influence benefit: $\Pi \gets \frac{1}{T} \sum_{i=1}^T \sum_{v \in A_i} b(v)$
    \For{each threshold $\tau \in \text{Thresholds}$}
        \State $\textsc{MotifProfit} \gets \textsc{ComputeMotifProfit}(\{A_i\}, \mathcal{M}, b, \tau, \mathcal{C}(\mathcal{S}))$
        \State Log result: Budget $\mathcal{B}$, Seed set $\mathcal{S}$, $\Pi$, MotifProfit, $\theta$, $\kappa$
    \EndFor
\EndFor
\end{algorithmic}
\end{algorithm}

The Reverse Influence Sampling (RIS) framework enhances classical influence maximization by incorporating motif-aware evaluation. It consists of three main parts: (i) estimation of KPT and generation of $\mathcal{RR}$ sets, (ii) greedy node selection for the seed set, and (iii) computation of motif-based profit. Algorithm~\ref{algorithm1:motif_ris} illustrates this process. Given a graph $\mathcal{G} = (\mathcal{V}, \mathcal{E}, \mathcal{P})$, node cost $\mathcal{C}(\cdot)$, benefit $b(\cdot)$, motif set $\mathcal{M}$, budgets $\mathbb{B}$, thresholds $\tau$, and simulation count $T$, the algorithm iteratively processes each budget $\mathcal{B}$. For each budget, it estimates the maximum number of seeds $k$ and computes the influence lower bound $\kappa$ using the \textsc{EstimateKPT} procedure (Line~4). This value determines the required number of reverse reachable sets $\theta$ (Line~5). $\mathcal{RR}$ sets are generated (\textsc{GenerateRRSets}, Line~6), and a greedy strategy then selects the seed set $\mathcal{S}$ under the budget constraint (Line~7). The diffusion process is simulated $T$ times to obtain the average profit (Lines~8--9). Finally, for each threshold $\tau$, motifs activated in the simulations are identified, their motif-based profit is computed (\textsc{ComputeMotifProfit}, Line~11), and results such as seed sets, influence profit, motif profit, and sampling parameters are recorded (Line~12).

\subsection{Part (i): Estimation of KPT and Generation of $\mathcal{RR}$ sets}
\begin{algorithm}[!htbp]
\caption{EstimateKPT}
\label{algorithm2:EstimateKPT}
\begin{algorithmic}[1]
\Statex \textbf{Input:} Graph $\mathcal{G}(\mathcal{V}, \mathcal{E} , \mathcal{P})$, Seed count $k$, Cost function $\mathcal{C}(\cdot)$, Benefit function $b(\cdot)$
\Statex \textbf{Output:} Estimated KPT value $\kappa$
\State $\eta \gets |\mathcal{V}|$, $m \gets |\mathcal{E}|$
\State Define $p_v \propto \frac{b(v)}{\mathcal{C}(v)}$ for all $v$
\For{$i = 1$ to $\log_2(n) - 1$}
    \State $c_i \gets$ Required samples at round $i$
    \State Initialize $sum \gets 0$
    \For{$j = 1$ to $c_i$}
        \State Sample node $v$ using $p_v$
        \State Generate $\mathcal{RR}(v)$ set 
        \State Estimate $\kappa_v = 1 - (1 - (\text{$|\mathcal{RR}(v)|$} / m))^k$
        \State $sum \gets sum + \kappa_v$
    \EndFor
    \If{$\frac{sum}{c_i} > \frac{1}{2^i}$}
        \State \Return $\kappa = \frac{n \cdot sum}{2 c_i}$
    \EndIf
\EndFor
\State \Return $\kappa = 1$
\end{algorithmic}
\end{algorithm}
The \textsc{EstimateKPT} procedure (Algorithm~\ref{algorithm2:EstimateKPT}) estimates the KPT value, a key parameter for determining the number of reverse reachable ($\mathcal{RR}$) sets needed in RIS-based algorithms. It first identifies the number of nodes $n = |V|$ and edges $m = |E|$ (Line~3). To emphasize nodes with higher influence, an importance sampling distribution is defined with probabilities proportional to $b(v)/\mathcal{C}(v)$ (Line~4). The estimation runs iteratively over logarithmic rounds $i = 1$ to $\log_2(n) - 1$ (Line~5). In each round, $c_i$ samples are drawn (Line~6), and for each sample a node $v$ is selected (Line~9), its $\mathcal{RR}$ set $\mathcal{RR}(v)$ is generated (Line~10), and the contribution $\kappa_v = 1 - (1 - |\mathcal{RR}(v)|/m)^k$ is computed (Line~11). The results are aggregated (Line~12), and the average $\tfrac{sum}{c_i}$ is compared with $1/2^i$ (Line~14). If satisfied, the algorithm returns $\kappa = \tfrac{n \cdot sum}{2c_i}$ (Line~15); otherwise, if no threshold is met in any round, it returns $\kappa = 1$ (Line~18).
\par \textbf{Complexity Analysis.} In Algorithm~\ref{algorithm2:EstimateKPT}, the main cost lies in the inner loop (Lines~6--10), where each of the $c_i$ samples requires generating an $\mathcal{RR}$ set. A single $\mathcal{RR}$ set generation may, in the worst case, traverse all $n = |\mathcal{V}|$ nodes and $m = |\mathcal{E}|$ edges. The outer loop (Line~3) executes for $\mathcal{O}(\log n)$ rounds, so the overall time complexity is $\mathcal{O}\!\left((n+m)\log n\right)$, as proved in \cite{NEURIPS2024_f50b83e4}. The space complexity is $\mathcal{O}(n+m)$, dominated by storing the graph and temporary $\mathcal{RR}$ sets.

\begin{algorithm}[!htbp]
\caption{GenerateRRsets}
\label{algorithm3:GenerateRRsets}
\begin{algorithmic}[1]
\Statex \textbf{Input:} $\theta$, $\mathcal{G}(\mathcal{V}, \mathcal{E}, \mathcal{P})$
\Statex \textbf{Output:} Set of $\mathcal{RR}$ sets $\mathcal{R}$
\State Sample $\theta$ start nodes using importance probabilities
\State For each node, perform reverse BFS
\State \Return $\mathcal{R}$
\end{algorithmic}
\end{algorithm}
The \textsc{GenerateRRsets} procedure (Algorithm~\ref{algorithm3:GenerateRRsets}) generates a collection of Reverse Reachable ($\mathcal{RR}$) sets, the core data structure in RIS-based influence maximization. The number of sets is given by $\theta = \frac{(8 + 2\epsilon)\, n \,(\,l \log n + \log \tbinom{n}{k} + \log 2\,)}{\text{KPT}\,\epsilon^{2}}$, with $\epsilon = 0.3$ and $l = 1$ as in \cite{NEURIPS2024_f50b83e4}. Given $\theta$, the algorithm samples nodes using an importance distribution proportional to their influence potential (Line~3). For each sampled node, a reverse BFS is performed (Line~4) to identify nodes that could reach it under the IC diffusion model. After all samples are processed, the complete set of $\theta$ many $\mathcal{RR}$ sets in $\mathcal{R}$ is returned (Line~5) for evaluating seed coverage in later steps.
\par \textbf{Complexity Analysis.} In Algorithm~\ref{algorithm3:GenerateRRsets}, each $\mathcal{RR}$ set is constructed by performing a reverse breadth-first search (BFS) on a randomly sampled live-edge graph under the IC model (Line~2). A single reverse BFS may, in the worst case, traverse all $n = |\mathcal{V}|$ nodes and $m = |\mathcal{E}|$ edges of the graph, leading to a cost of $O(n+m)$ per $\mathcal{RR}$ set. Since the algorithm generates $\theta$ many $\mathcal{RR}$ sets, the total worst-case time complexity is $O(\theta (n+m))$. The space complexity is $\mathcal{O}(\theta \cdot |\mathcal{RR}|)$ = $\mathcal{O}(\theta \cdot n)$ (when in the worst case, the size of an $\mathcal{RR}$ set is $n$), for storing all generated $\mathcal{RR}$ sets.
\subsection{Part (ii): Greedy-Based approach for node selection of the seed set}
\vspace{-0.2in}
\begin{algorithm}[!htbp]
\caption{GreedySeedSelection}
\label{algorithm4:GreedySeedSelection}
\begin{algorithmic}[1]
\Statex \textbf{Input:} $\mathcal{RR}$ sets $\mathcal{R}$, Cost function $\mathcal{C}(\cdot)$, Benefit function $b(\cdot)$, Budget $\mathcal{B}$
\Statex \textbf{Output:} Seed set $\mathcal{S}$
\State Initialize $\mathcal{S} \gets \emptyset$, RemainingBudget $\gets \mathcal{B}$
\While{RemainingBudget $> 0$}
    \ForAll{nodes $v \notin \mathcal{S}$}
        \State Compute coverage score: number of $\mathcal{RR}$ sets containing $v$
        \State Compute normalized score: $\text{score} / \mathcal{C}(v)$
    \EndFor
    \State Select node $v^*$ with highest normalized score within budget
    \If{no such node exists}
        \State \textbf{break}
    \EndIf
    \State $\mathcal{S} \gets \mathcal{S} \cup \{v^*\}$,
           RemainingBudget $\gets$ RemainingBudget - $\mathcal{C}(v^*)$
    \State Mark RR sets covered by $v^*$
\EndWhile
\State \Return $\mathcal{S}$
\end{algorithmic}
\end{algorithm}
The \textsc{GreedySeedSelection} procedure (Algorithm~\ref{algorithm4:GreedySeedSelection}) selects an optimal seed set under a budget constraint using precomputed $\mathcal{RR}$ sets. It starts with an empty seed set $\mathcal{S}$ and budget $\mathcal{B}$ (Line~1), then iteratively adds nodes while the budget remains (Line~2). In each round, all candidate nodes $v \notin \mathcal{S}$ are evaluated: their coverage score is the number of $\mathcal{RR}$ sets containing $v$ (Line~4), normalized by cost $\mathcal{C}(v)$ (Line~5). The node $v^{*}$ with the highest affordable normalized score is chosen (Line~7). If no node fits the budget, the loop terminates (Lines~8--10). Otherwise, $v^{*}$ is added to $\mathcal{S}$, its cost subtracted, and its covered $\mathcal{RR}$ sets marked (Lines~11--12) to avoid double counting. This process repeats until the budget is exhausted or no nodes remain, and the final seed set $\mathcal{S}$ is returned (Line~14).
\par \textbf{Complexity Analysis.} In Algorithm~\ref{algorithm4:GreedySeedSelection}, initialization (Line~1) takes $\mathcal{O}(1)$ time. The main cost is the \texttt{while} loop (Lines~2--13), which may run up to $\mathcal{B}/C_{\min}$ times, where $C_{\min} = \min_{v \in \mathcal{V}} \mathcal{C}(v)$. In each iteration, computing coverage scores for all $n = |\mathcal{V}|$ nodes requires processing $\theta$ $\mathcal{RR}$ sets, giving $\mathcal{O}(n \cdot \theta \cdot |\mathcal{RR}(v)|)$. In the worst case $|\mathcal{RR}(v)| = n$, i.e., $\mathcal{O}(\theta \cdot n^2 )$. Normalized score computation (Line~5) and node selection (Line~7) add $\mathcal{O}(n)$ each, but are dominated as well. Marking covered $\mathcal{RR}$ sets (Line~12) takes $\mathcal{O}(\theta \cdot n)$. Hence, each iteration costs $\mathcal{O}(\theta \cdot n^2)$, and the overall time complexity is $\mathcal{O}\!\left(\tfrac{\mathcal{B}}{C_{\min}} \cdot \theta \cdot n^2 \right)$. The space complexity is $\mathcal{O}(n + \theta \cdot n)$ = $\mathcal{O}(\theta \cdot n)$ (as $\mathcal{RR}$ sets dominate the overall cost), for maintaining node scores and marked $\mathcal{RR}$ sets.
\subsection{Part (iii): Computation of Motif-based Profit}
\vspace{-0.2in}
\begin{algorithm}[!htbp]
\caption{ProcessMotifProfit}
\label{algorithm5:ProcessMotifProfit}
\begin{algorithmic}[1]
\Statex \textbf{Input:} Simulation results $\{A_i\}$, Motifs $\mathcal{M}$, Benefit $b(\cdot)$, Threshold $\tau$, Seed cost $\mathcal{C}(\mathcal{S})$
\Statex \textbf{Output:} Average motif-based profit
\For{each simulation $A_i$}
    \State Identify $\mathcal{M}_i \gets \{m \in \mathcal{M} \mid |m \cap A_i| \geq \tau\}$
    \State $B_i \gets \sum_{v \in \cup \mathcal{M}_i} b(v)$
    \State $\Pi_i \gets B_i - \mathcal{C}(\mathcal{S})$
\EndFor
\State \Return $\dfrac{1}{T} \sum_i \Pi_i$
\end{algorithmic}
\end{algorithm}
The \textsc{ProcessMotifProfit} procedure (Algorithm~\ref{algorithm5:ProcessMotifProfit}) concludes the RIS framework by computing the average motif-based profit from a given seed set using multiple diffusion simulations under the IC model. It takes as input the simulation results $\{A_i\}$, a motif set $\mathcal{M}$, benefit function $b(\cdot)$, threshold $\tau$, and seed cost $C$. For each simulation $A_i$, motifs $\mathcal{M}_i$ are considered active if $|m \cap A_i| \geq \tau$ (Line~1--2). The benefit $B_i$ is then the sum of $b(v)$ over all nodes in the union of active motifs (Line~3), and the net profit $\Pi_i$ is $B_i - \mathcal{C}(\mathcal{S})$ (Line~4). After $T$ simulations, the average profit $\tfrac{1}{T}\sum_i \Pi_i$ is returned (Line~6).
\par \textbf{Complexity Analysis.} The time complexity of Algorithm~\ref{algorithm5:ProcessMotifProfit} depends on the number of simulations $T$, motifs $|\mathcal{M}|$, motif size $s$, and nodes $n$. In each simulation, checking threshold activation costs $\mathcal{O}(s)$ per motif, or $\mathcal{O}(|\mathcal{M}|\,s)$ in total. Computing the benefit $B_i$ then requires at most $\mathcal{O}(n)$ time. Hence, the per-simulation cost is $\mathcal{O}(|\mathcal{M}|\,s + n)$, and over $T$ simulations the total runtime is $\mathcal{O}\!\left(T(|\mathcal{M}|\,s + n)\right)$. The space complexity is $\mathcal{O}(|\mathcal{M}|s + n)$, for storing motif definitions and activated nodes.

Therefore, the overall time complexity of our proposed approach (Algorithm~\ref{algorithm1:motif_ris}) is $\mathcal{O}\!\Bigg((n+m)\log n \;+\; \theta (n+m) \;+\; \tfrac{\mathcal{B}}{C_{\min}} \cdot \theta \cdot n^{2} \;+\; T(|\mathcal{M}|\,s+n) \Bigg)$. The overall space complexity is $\mathcal{O}(n+m+\theta \cdot n+|\mathcal{M}|\,s+T \cdot n)$.

\section{Experimental Evaluation} \label{Sec:Experiments}
We next present the experimental evaluation of the proposed approach, beginning with the datasets.
\subsection{Dataset Description}
Our experiments use the following networks:
\begin{itemize}
    \item \textbf{US Congress} (Congress) \cite{fink2023centrality,fink2023twitter}: Twitter interaction network for the 117th United States Congress (House and Senate).
    \item \textbf{Email-Eu-Core} (Euemail) \cite{yin2017local, leskovec2007graph}: Built from email exchanges in a large European research institution; an edge $(u,v)$ exists if $u$ sent $v$ at least one email. 
    \item \textbf{Wikipedia Vote} (Wikivote) \cite{leskovec2010signed,leskovec2010predicting}: Voting data from Wikipedia’s inception to Jan 2008; nodes are users and a directed edge $(u,v)$ means $u$ voted on $v$.
\end{itemize}
\vspace{-0.2in}
\begin{table*}[!htbp]
    \centering
    \begin{tabular}{|l|c|c|c|c|c|}
        \hline
        \textbf{Dataset Name} &
        \makecell{\textbf{Type of}\\\textbf{Graph}} &
        \makecell{\textbf{Number of}\\ \textbf{Nodes}\\$|\mathcal{V}|$} &
        \makecell{\textbf{Number of}\\ \textbf{Edges}\\$|\mathcal{E}|$} &
        \makecell{\textbf{Maximum}\\ \textbf{Degree}\\$d_{\max}$} &
        \makecell{\textbf{Average}\\ \textbf{Degree}\\$d_{\text{avg}}$} \\
        \hline
        US Congress    & Directed      & 475   & 13289  & 284   & 55.95    \\ \hline
        Email-Eu-Core  & Directed      & 1005  & 25571  & 546     & 50.89  \\ \hline
        Wiki-Vote      & Directed      & 7115  & 103689 &  1167  &  29.15  \\ 
        \hline
    \end{tabular}
    \caption{Basic statistics of the datasets used in our experiments.}
    \label{1Tab:datasets}
\end{table*}
\vspace{-0.2in}
All the datasets have been downloaded from Stanford Large Network Dataset Collection \footnote{\url{https://snap.stanford.edu/data/index.html}}. Table $\ref{1Tab:datasets}$ describes the basic statistics of the datasets.

\subsection{Experimental Setup}
In our study, several parameters need to be defined, beginning with the influence probability setting.
\paragraph{Influence Probability}: We consider two settings:
\begin{itemize}
    \item \textbf{Trivalency}: Each edge is assigned a probability uniformly at random from $\{0.1,0.01,0.001\}$.
    \item \textbf{Weighted Cascade}: Each edge $(u,v)$ has probability inversely proportional to the in-degree of $v$, i.e., $\frac{1}{deg^{in}(v)}$. 
\end{itemize}
\paragraph{Cost and Benefit Values}: We use a degree-proportional cost setting, where a higher out-degree implies a higher cost, as in practice (e.g., celebrities with more followers demand higher fees). The benefit of each node is then assigned by scaling its cost.

\subsection{Baseline Methods}
We compare our proposed methodology against the following baselines:
\begin{itemize}
    \item \textbf{Random}: Nodes are selected randomly until the budget is exhausted.
    \item \textbf{High Degree}: Nodes are ranked by degree and selected in order until the budget is exhausted.
    \item \textbf{CELF}: A popular influence maximization algorithm by Leskovec et al. \cite{leskovec2007cost}, adapted here for profit maximization. 
    \item \textbf{Simple Greedy}: Starting with an empty set, nodes are added iteratively based on marginal profit gain \cite{kempe2003maximizing}.
\end{itemize}
All baselines were implemented in Python 3.0.1+ with NetworkX 2.2.1, and experiments were run on a Linux desktop with 64 GB RAM and a 32-core Intel i9 processor.

\subsection{Experimental Results and Discussions}
\par
\begin{figure}[!htbp]
  \centering
  \includegraphics[width=\linewidth]{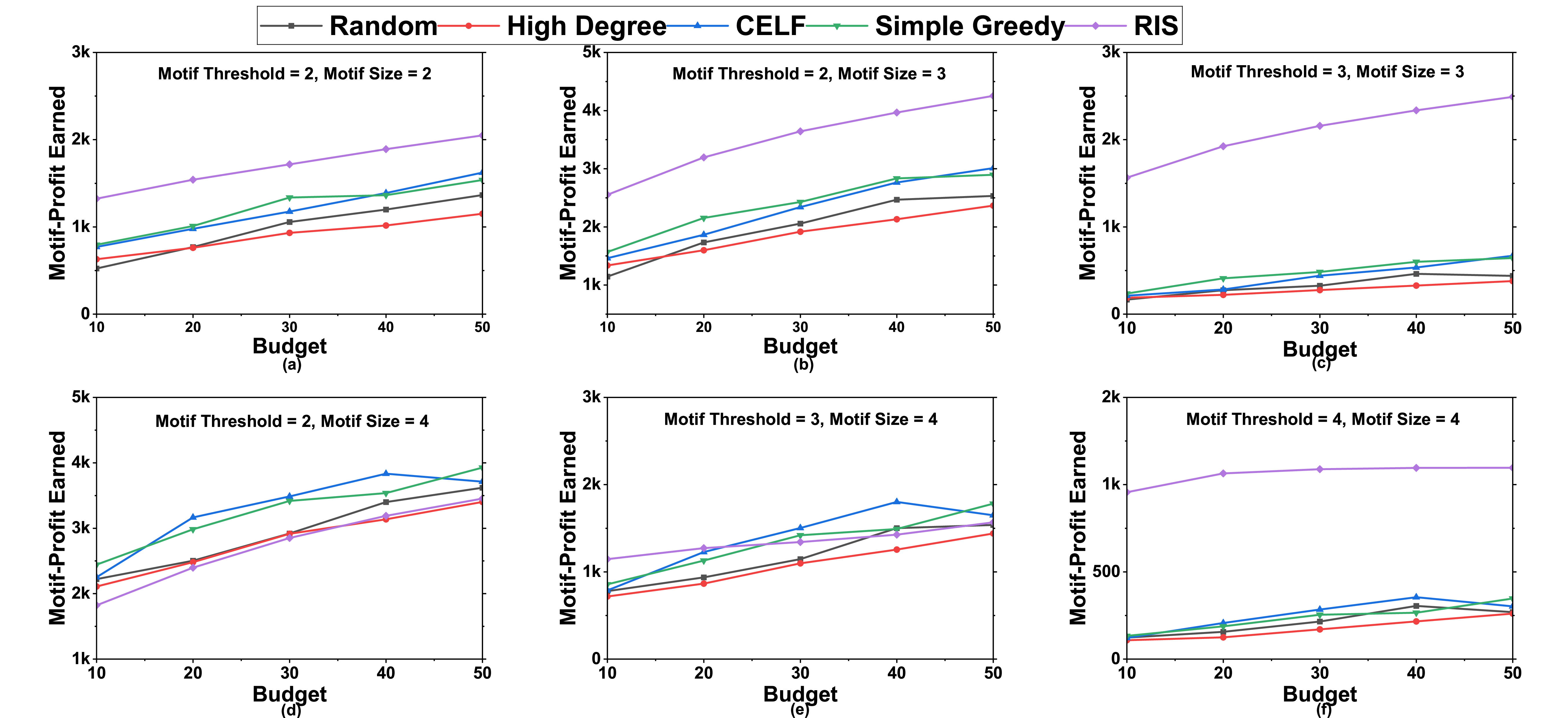}
  \caption{Budget vs. Motif-Profit for Congress (Trivalency).}
  \label{1Fig:CongressTrivalencyProfitEarned}
\end{figure}

\begin{figure}[!htbp]
  \centering
  \includegraphics[width=\linewidth]{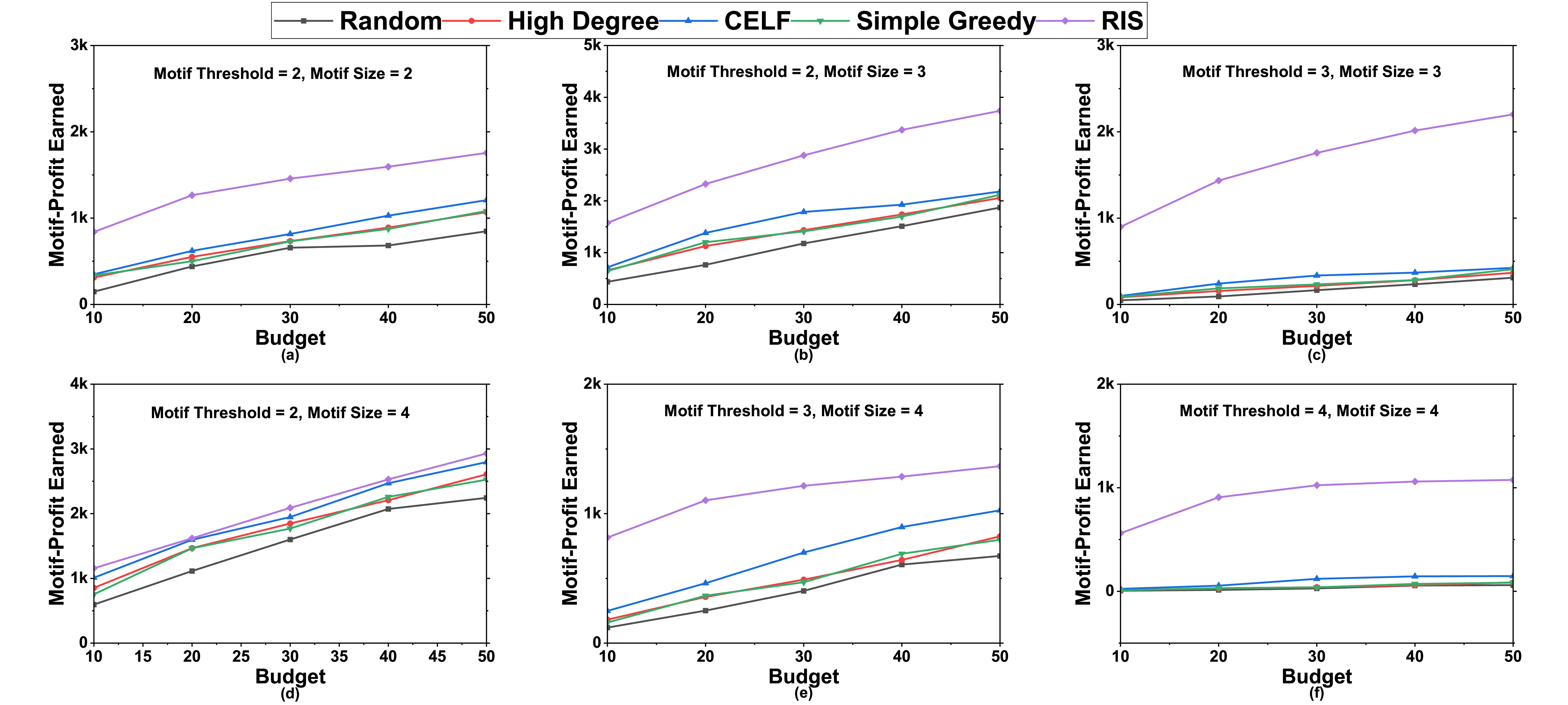}
  \caption{Budget vs. Motif-Profit for Congress (Weighted Cascade).}
  \label{2Fig:CongressWCProfitEarned}
\end{figure}

\begin{figure}[!htbp]
  \centering
  \includegraphics[width=\linewidth]{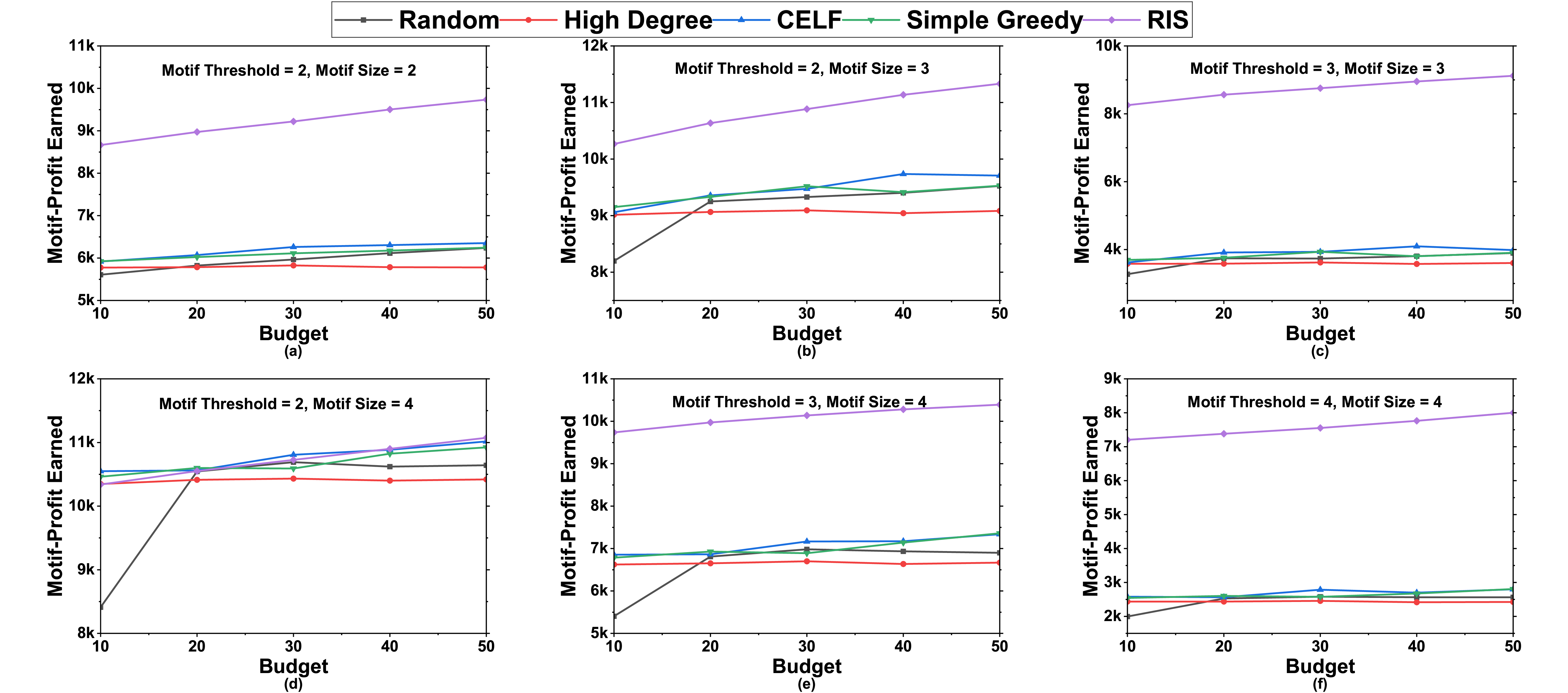}
  \caption{Budget vs. Motif-Profit for Euemail (Trivalency).}
  \label{3Fig:EuemailTrivalencyProfitEarned}
\end{figure}

\begin{figure}[!htbp]
  \centering
  \includegraphics[width=\linewidth]{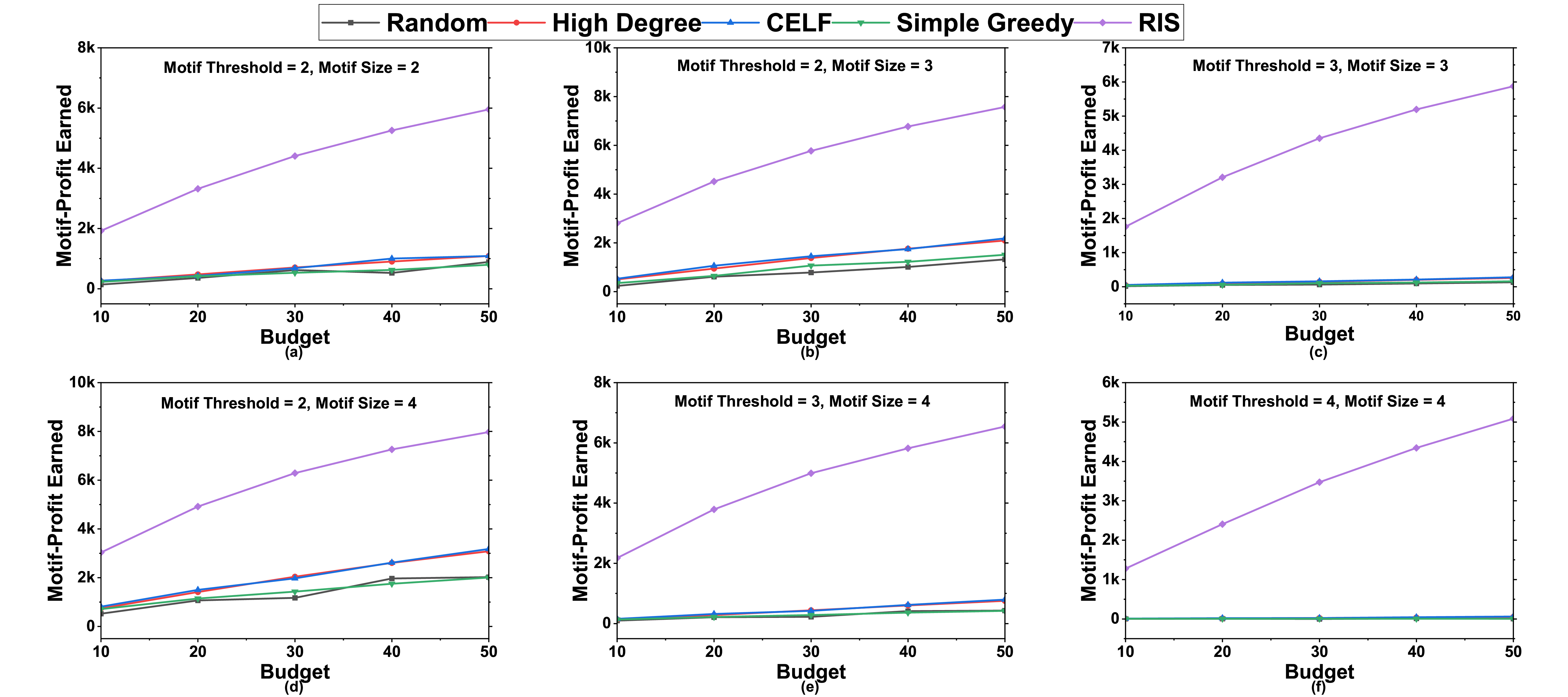}
  \caption{Budget vs. Motif-Profit for Euemail (Weighted Cascade).}
  \label{4Fig:EuemailWCProfitEarned}
\end{figure}

\begin{figure}[!htbp]
  \centering
  \includegraphics[width=\linewidth]{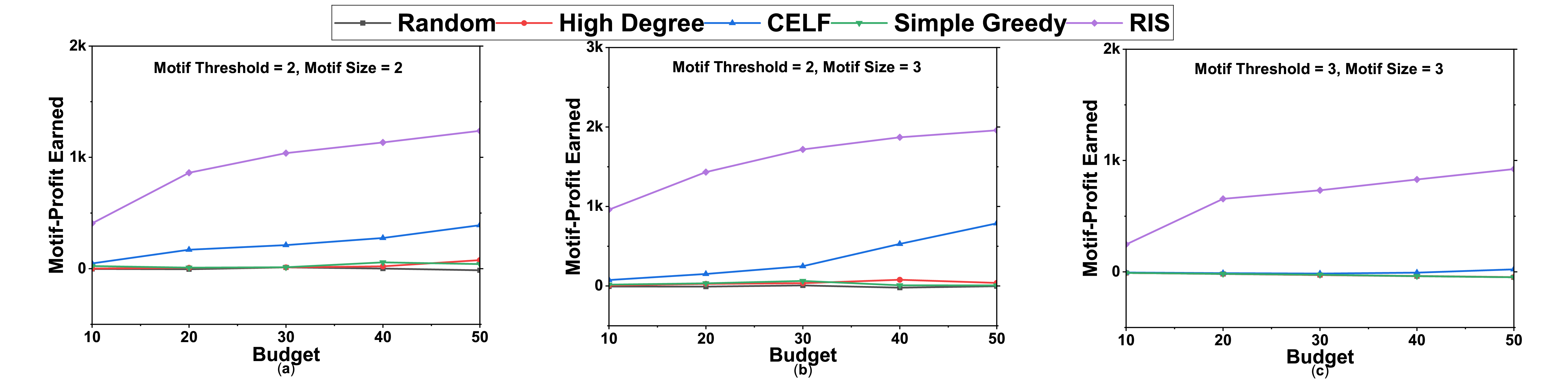}
  \caption{Budget vs. Motif-Profit for Wikivote (Weighted Cascade).}
  \label{5Fig:WikivoteWCProfitEarned}
\end{figure}

 We have analyzed three datasets for our experiments, which are listed in Table~\ref{1Tab:datasets}. Our research objective is to evaluate how well the algorithms work with the structural pattern of the graph, i.e., motifs. The metric used for comparison is the motif-based profit earned under a given budget. We also aim to understand which threshold values are effective in maximizing motif-based profit. For the Congress dataset (Figures~\ref{1Fig:CongressTrivalencyProfitEarned} and \ref{2Fig:CongressWCProfitEarned}), experiments were conducted for motif sizes 2, 3, and 4. We compared Random, High Degree, CELF, and Simple Greedy with the Motif Oriented RIS approach. Across all probability settings, motif profit increases monotonically with budget, and RIS consistently outperforms all other algorithms. Simple Greedy and CELF are often competitive, while Random and High Degree perform similarly. For instance, at budget 10 under Trivalency probability settings, RIS achieves 1323.8 units of profit, compared to 794.9 for Simple Greedy (66\% less) and 770.5 for CELF (71\% less). High Degree and Random perform much worse, with RIS producing 110\% and 153\% higher profits, respectively. Similar trends are observed across other budgets and motif sizes. In the Weighted Cascade probability setting, RIS again outperforms all approaches by large margins (Figures~\ref{2Fig:CongressWCProfitEarned} (a)–(f)). On average, RIS generates about 10990\% more profit than Random, 4644\% more than High Degree, 2222\% more than CELF, and 5873\% more than Simple Greedy. The Euemail dataset shows even more striking results (Figures~\ref{3Fig:EuemailTrivalencyProfitEarned}, \ref{4Fig:EuemailWCProfitEarned}). Under Trivalency, RIS outperforms by 260\% over Random, 229\% over High Degree, 188\% over CELF, and 193\% over Simple Greedy. Under Weighted Cascade, RIS gains are enormous, reaching up to 85586\% compared to Random, 22509\% over High Degree, 16105\% over CELF, and 81017\% over Simple Greedy. The Wikivote dataset (Figure~\ref{5Fig:WikivoteWCProfitEarned}) further confirms this trend. Random, High Degree, and Simple Greedy perform about 105\% worse than RIS, while CELF is 102\% worse.

We next analyze the impact of threshold values of different motif sizes. In the Congress dataset, at threshold 2, motif profits are higher as motifs are activated more easily. RIS remains the best performer, e.g., achieving 1755 units at budget 50 under Weighted Cascade probability setting for motif size 2 (Figure~\ref{2Fig:CongressWCProfitEarned} (a)). CELF and Simple Greedy reach about 60–100\% of RIS’s performance, while Random and High Degree remain significantly lower. At threshold 3, profits decline for all algorithms, though RIS still leads. Stricter thresholds reduce overall profits since fewer motifs activate, but RIS continues to dominate. For example, at budget 40 under Weighted Cascade, RIS profits are reduced but still higher than all baselines, with Random and High Degree nearly negligible. In Figure~\ref{1Fig:CongressTrivalencyProfitEarned}, CELF achieves up to 79\% of RIS at threshold 2 for motif size 2 and remains closer to RIS in motif sizes 3 and 4 for thresholds 2 and 3. The Euemail dataset shows similar threshold effects. Lower thresholds yield higher profits, while values closer to motif size reduce profits. For example, in Trivalency with motif size 4 at budget 10 (Figure~\ref{3Fig:EuemailTrivalencyProfitEarned} (d)–(f)), Random achieves 8410.31 at threshold 2, 5403.61 at threshold 3, and only 1997.59 at threshold 4. RIS consistently leads: in Weighted Cascade with motif size 4 at budget 10 (Figure~\ref{4Fig:EuemailWCProfitEarned} (d)–(f)), it performs 326\% better than Simple Greedy at threshold 2, 1506\% better at threshold 3, and 20483\% better at threshold 4. The Wikivote dataset also reflects these effects (Figure~\ref{5Fig:WikivoteWCProfitEarned} (b)–(c)): at budget 30 for motif size 3, Simple Greedy performs 97\% worse (threshold 2) and 103\% worse (threshold 3) than RIS.  

Across all datasets, RIS consistently outperforms Random, High Degree, CELF, and Simple Greedy, confirming its superiority in exploiting motifs. While CELF and Simple Greedy are occasionally competitive, Random and High Degree perform poorly, especially under the Weighted Cascade probability setting. Threshold analysis shows that smaller thresholds give higher profits, while stricter ones reduce them across all algorithms. Nonetheless, RIS remains the clear leader, showing robustness under varying activation criteria.

% In preamble (helps LaTeX pack floats better)
\renewcommand{\topfraction}{0.95}
\renewcommand{\bottomfraction}{0.95}
\renewcommand{\textfraction}{0.05}
\renewcommand{\floatpagefraction}{0.9}
\renewcommand{\dbltopfraction}{0.95}
\renewcommand{\dblfloatpagefraction}{0.9}

% Congress (Trivalency)

% Congress (WC)

% Euemail (Trivalency)

% Euemail (WC)

% Wikivote (WC)

\section{Concluding Remarks} \label{Sec:Conclusion}
In this paper, we have studied the Motif-Oriented Profit Maximization Problem, where, given a social network with the selection cost of the nodes and a set of motifs with their corresponding benefit value, this problem asks to choose a limited number of highly influential nodes within a budget such that the motif oriented earned profit gets maximized. This problem is NP-hard to solve optimally. We have proposed a reverse reachable set-based solution approach. The experimental results with real-world social network datasets show the effectiveness of the proposed solution approach, motif oriented RIS. Our approach proves to be robust across datasets, probability settings, and threshold values, establishing its effectiveness in exploiting motif structures for profit maximization. Now, our future study on this problem will remain concentrated on developing more efficient solution methodologies.
\newpage
\bibliographystyle{splncs04}
\bibliography{Paper}

\end{document}